# Tuna-step: Tunable Parallelized Step Emulsification for the Generation of Droplets with Dynamic Volume Control to 3D Print Functionally Graded Porous Materials

Francesco Nalin[1], Maria Celeste Tirelli[1], Piotr Garstecki[1], Witold Postek[1,2]*, Marco Costantini[1]*

[1]   Institute of Physical Chemistry, Polish Academy of Sciences, 44/52 ul. Kasprzaka, 01-224 Warsaw, Poland

[2]   Broad Institute of MIT and Harvard, Merkin Building, 415 Main St, Cambridge, MA 02142, USA.

\*   **Correspondence:** mcostantini@ichf.edu.pl; wpostek@broadinstitute.org

## Abstract

We present Tuna-step, a novel microfluidic module based on step emulsification that allows for reliable generation of droplets of different sizes. Until now, sizes of droplets generated with step emulsification were hard-wired into the geometry of the step emulsification nozzle. To overcome this, we incorporate a thin membrane underneath the step nozzle that can be actuated by pressure, enabling the tuning of the nozzle size on-demand. By controllably reducing the height of the nozzle, we successfully achieved a three-order-of-magnitude variation in droplet volume without any adjustment of the flow rates of the two phases. We further enhanced our system by developing and applying a new hydrophilic surface modification, ensuring long-term stability and preventing swelling of the device when generating Oil-in-Water droplets. We used our system in the manufacturing of functionally graded soft materials with adjustable porosity and material content. By combining our microfluidic device with a custom 3D printer, we generated and extruded Oil-in-Water emulsions in an agarose gel bath, allowing for the creation of unique self-standing 3D hydrogel structures with porosity decoupled from flow rate and with composition gradients of external phases. We upscaled Tuna-step by setting 14 actuatable nozzles in parallel, offering a step-emulsification-based single chip solution that can accommodate various requirements in terms of throughput, droplet volumes, flow rates, and surface chemistry.

## 1. Introduction

Microfluidic droplet generation technologies enable the formation of small droplets – ranging from femtolitres to nanolitres – by emulsifying immiscible fluids. Since their introduction, microfluidic emulsifiers have found widespread applications in various fields, including food production, advanced material synthesis, and biochemical research[1–4]. Among the available platforms, step emulsification[5,6] has lately emerged as a simple, yet versatile system for droplet production. This process involves driving a liquid through a narrow constriction that terminates in a wider chamber filled with a second immiscible phase. The significant difference in capillary pressure between the nozzle and the chamber causes the liquid to break into droplets. This method produces monodisperse droplets with volumes determined by the orifice size, as long as the generated droplets escape the vicinity of the step before new droplets start to form. This feature, combined with relative ease of parallelization of step emulsification nozzles, makes step emulsification attractive for high-throughput applications[7–9] and ensures uniform and geometry-embedded droplet size within a safe range of flow rates[10], unlike other methods such as cross-flow[11,12], co-flow[13,14] or active control[15].

While step emulsification devices offer robustness against changes in flow rate, the inability to manipulate droplet volumes after chip manufacturing poses a limitation for their use in domains that necessitate varying sizes while maintaining high throughput, such as microfluidic-assisted synthesis of porous materials. Recently, emulsions generated using microfluidics have been demonstrated to be a suitable template for the synthesis of porous materials with controlled morphologies, in particular of materials characterized by monodisperse pores. Moreover, in a very recent study by Marcotulli *et al*.[16], it has been demonstrated that one can synthesize soft porous functionally graded hydrogels with spatially controlled chemical composition and porous architecture by combining conventional microfluidic droplet generators and 3D printing technologies.

The impact of such microfluidic-assisted 3D printing strategies in the field of porous functionally graded materials (pFGMs) synthesis may be enormous. The possibility to have either compositional, micro-architectural or both kinds of gradients within a single material has the potential to generate a virtually limitless number of new materials with tailored and unprecedented structural and functional properties. However, despite the promising pioneering work[16–19], the field is still at its infancy and many technical challenges must be addressed before these systems can be translated into actual advanced solutions. In particular, the production throughput and the precise, on-demand control over droplet size, volume fraction and spatial position still represent major hurdles.

To address these limits, we have developed Tuna-step, a PDMS-based microfluidic system that enables on-demand control over the size of the step emulsification nozzle orifice. This is achieved by pressure-actuating a thin membrane positioned beneath the emulsifying nozzle that dynamically changes the nozzle geometry. Notably, we demonstrate how this feature translates into the decoupling of the flow rates of the two immiscible phases from the emulsion features, such as droplet size and volume fraction. Here, we first analysed how Tuna-step can be effectively used both for the production of W/O and O/W emulsions, the latter being possible thanks to an innovative hydrophilic PDMS surface modification strategy that we additionally developed. Following that, we integrated Tuna-step within a custom 3D printing platform and extruded O/W emulsions into a granular agarose fluid–gel bath. This strategy allows to further separate the rheological properties of the emulsion ink from its printability and achieve precise spatial positioning during the extrusion process. Our versatile design enabled us to additionally achieve multi-material 3D deposition by rapidly switching between different continuous phases. Finally, we demonstrated the scalability potential of our Tuna-step by producing droplets using a 14-nozzle device, increasing the system throughput by a factor of ~14, an aspect particularly important for the manufacturing of macroscopic pFGMs.

## 2. Results

**2.1  Design of the microfluidic Tuna-step system and its working principle**

The microfluidic tuna-step system consists of a two-layer PDMS chip - as two layers are needed for an on-chip microvalve to function[20–22] - bonded to a common glass slide (**Figure 1a-c**). The top layer, referred to as the thick layer, comprises a narrow channel that serves as the inlet for the dispersed phase. This channel directs the liquid to a spacious emulsification chamber. Additionally, two side channels enable a continuous flow of the continuous phases from the sides of the nozzle. The emulsification channel culminates in the chip outlet, facilitating the collection of the emulsion.

The middle layer, known as the thin layer, features a square chamber situated below the step-emulsification nozzle. A thin membrane, approximately 50 µm in thickness, separates this chamber from the nozzle. When pressure is applied to the chamber, the thin membrane undergoes deflection, causing a reduction in the size of the nozzle (**Figure 1e, f**). To better investigate the effect of membrane deflection over the step size, we introduced a water solution of a fluorescently-labelled polymer (gelatin-FITC) into the chip and conducted a fluorescence scan along the chamber using a confocal microscope (**Figure 1f**). As expected, the membrane deflects, resulting in a significant decrease in the nozzle size. Once the applied pressure exceeds 50 mbar, the thin membrane, which in the central part of the step deforms the most, comes in contact with the opposite nozzle wall. Under these conditions, two streams of liquid emerge at the sides of the step, we call those streams side nozzles. As the pressure continues to increase, the cross-sectional area of the two side nozzles decreases. When the pressure is released, the elastic membrane returns to its initial configuration.

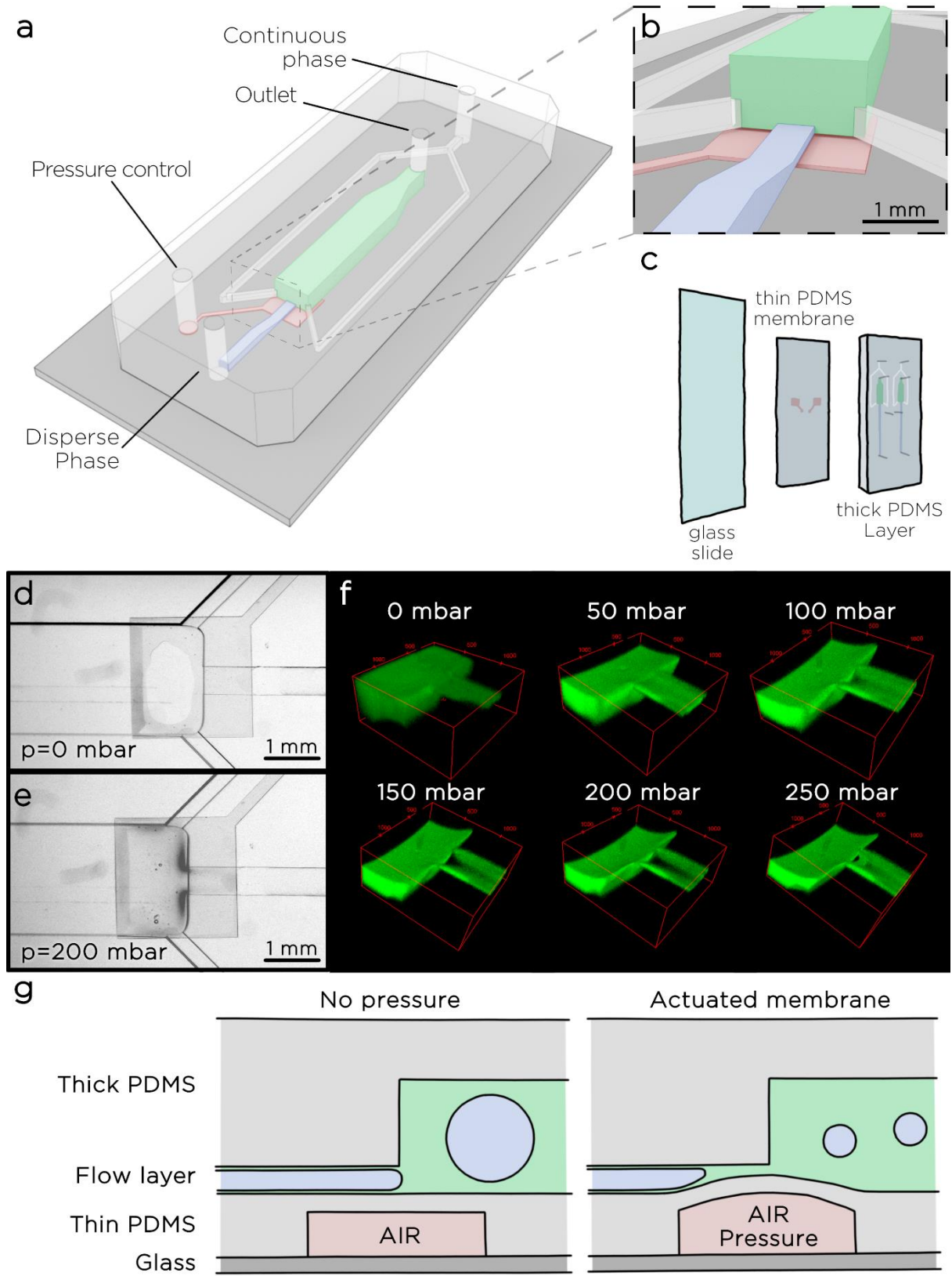

**Figure 1. Scheme of the chip** – *a)* Graphical representation of the chip including one narrow channel for the disperse phase (light blue), ending on the step with the emulsification chamber (green). Two side channels (white) allow constant refill of the continuous phase and help the flow from the droplet-generating step to the outlet of the chip. On a separate layer, the pressure-regulated membrane (red).

*b)* Zoom on the step. *c)* Scheme of the different layers of the chip. *d, e)* Two images of the same chip before and after membrane actuation. *f)* Confocal images of the nozzle filled with fluorescent molecules. During the imaging, pressurized air was applied to the membrane layer, promoting the increasing deflection of the membrane. *g)* Scheme of the step before and after the actuation of the membrane. After actuation, the reduction of the height of the nozzle results in the decrease of the droplet size.

## 2.2      Characterization of Tuna-step for the generation of Water-in-Oil emulsions

As initial experiments, we measured the diameter of a series of water droplets generated in NOVEC 7500 oil containing 2% w/v surfactant in a single nozzle chip (**Figure 2**). Both phases were supplied to the microfluidic system using precision syringe pumps while the membrane was actuated using a pressure controller. We fixed the flow rate for the oil at 1 µl/min and set the flow rate for the aqueous phase at 0.5, 1, 2, 3, and 5 µl/min. For each set of flow rates, we monitored the effects of the actuatable membrane on droplet generation for a range of pressures ($P_m$) starting from 0 mbar and increasing with 10 mbar steps. In each experiment, $P_m$ was increased until the system stopped operating smoothly. The maximum pressure applied to the membrane depended on the droplet flow rate used, with an inversely proportional trend between the two variables: the higher the droplet flow rate, the lower the maximum $P_m$ that can be used to generate droplet in a stable and reliable manner.

For $P_m$ = 0 mbar, as expected, the generated droplets reached their maximum volume as the step size was at a maximum. Interestingly, we observed that during droplet generation the membrane was partially deflected downward and, in some cases depending on the droplet flow rate used, touched the glass slide. Nevertheless, this aspect did not affect droplet generation. At $P_m$ = 10 mbar, the applied pressure lacks the necessary strength to consistently deform the membrane and the droplets formed at this pressure have inconsistent sizes (see larger error bars in **Figure 2b**). At 20 mbar and higher pressures, the droplet generation became reliable, with an increase of $P_m$ resulting in a consistent decrease in nozzle and droplet sizes (**Figure 2b**). Based on the explored operational ranges, we obtained a dynamic control over droplet size between 64±4 µm and 551±2 µm corresponding to a volume range between 140±30 pl and 88±1 nl.

7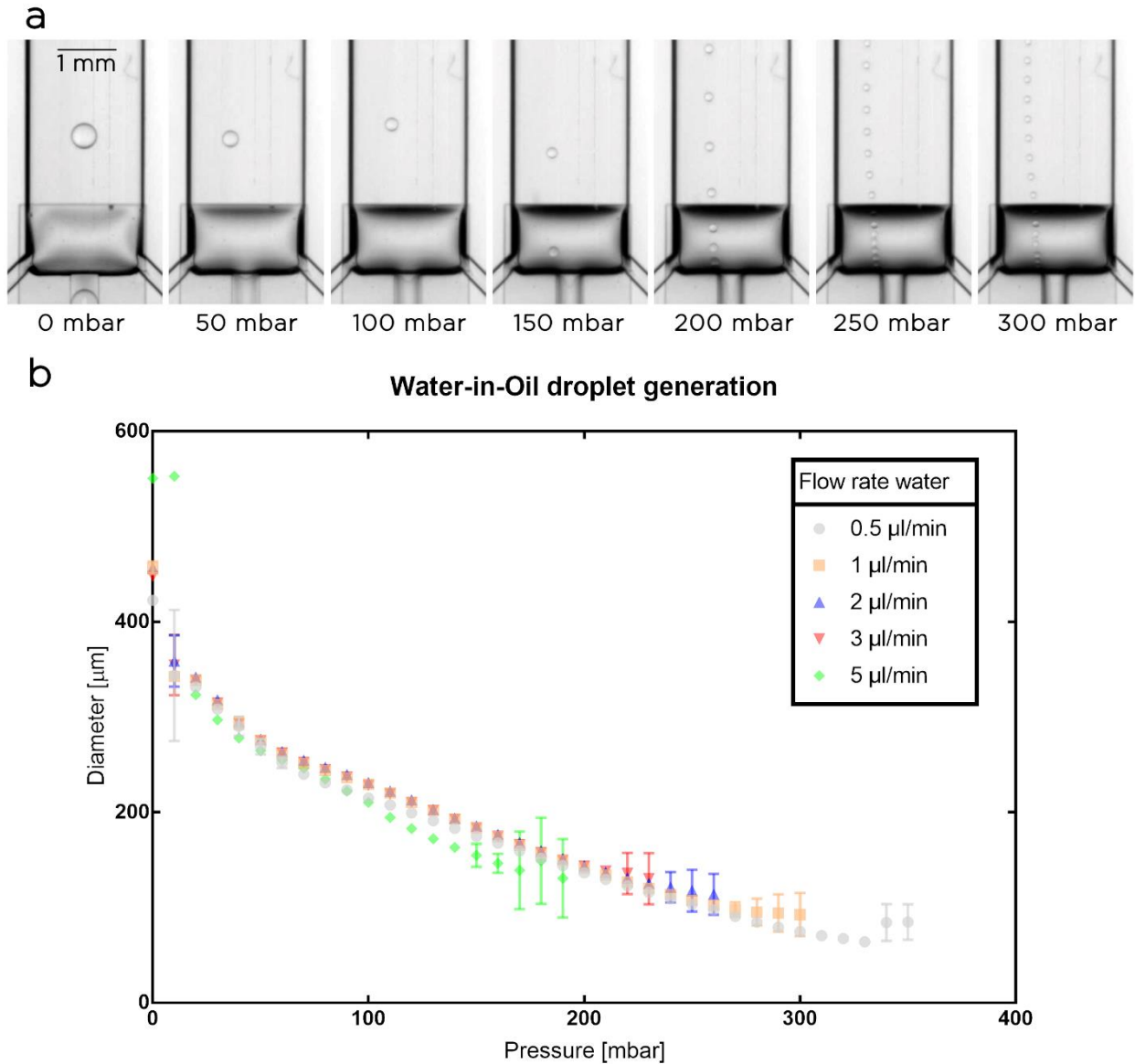

**Figure 2**. **Tuna-step calibration for Water-in-Oil droplets** – *a) Generation of water droplets in NOVEC 7500 oil at a constant flow rate (Water 0.5 µl/min, Oil 1 µl/min) with increasing pressure on the membrane layer (from 0 to 300 mbar). b) Calibration of droplet diameter as a function of $P_m$ for different water flow rates and a constant oil flow rate (1 µl/min). For each experimental set of flow rates, the maximum applicable $P_m$ was determined; above this pressure droplet generation becomes unstable.*

## 2.3   PDMS Hydrophilic surface modification for prolonged Oil-in-Water emulsion generation

Following the preliminary characterization, we decided to explore the possibility to produce O/W emulsions with Tuna-step chips. This required satisfying two essential conditions: i) the inner surface of the microfluidic channels must be hydrophilic to ensure the proper sliding of the oil droplets generated after the step, and ii) the chip must be impermeable towards the oil phase (most of the organic solvents and oils quickly swell PDMS)[23].

To address these two requirements, we developed an innovative surface modification that promotes both a hydrophilic surface and prevents the PDMS swelling due to the oil phase, enabling long-time emulsion production. The surface modification consists of two simple steps, realized directly after the standard plasma bonding[24] of the chip (**Figure 3a)**: i) coating the microchannels with vinyltrichlorosilane and ii) flushing, under UV light, the treated chip with a Gelatin-methacryloyl (GelMA) water solution containing a photoinitiator to chemically bond GelMA polymer chains to the vinyl moieties previously introduced onto PDMS surface.

To confirm the validity and the stability of our surface modification, we measured both the water contact angle and the generation of O/W emulsions after each step of the protocol using a standard T-junction (**Figure 3c**, after plasma bonding, after silanization and after GelMA treatment). As shown in **Figure 3c** the contact angle of untreated PDMS was 117°; after silanization with vinyltrichlorosilane, the value decreased to 108°, while the one of the GelMA-treated PDMS was 71°, confirming the hydrophilicity of the PDMS surface. To generate O/W emulsions, we used hexadecane as droplet phase and recorded time-lapse videos over a period of up to 24h. While the droplet generation remained stable in the surface-modified chip for the whole studied period (24 consecutive hours of emulsion generation), the other two chips (plasma bonded only, plasma bonded and silanized) presented visible swelling of the PDMS channels already after 30 minutes (**Figure 3d**).

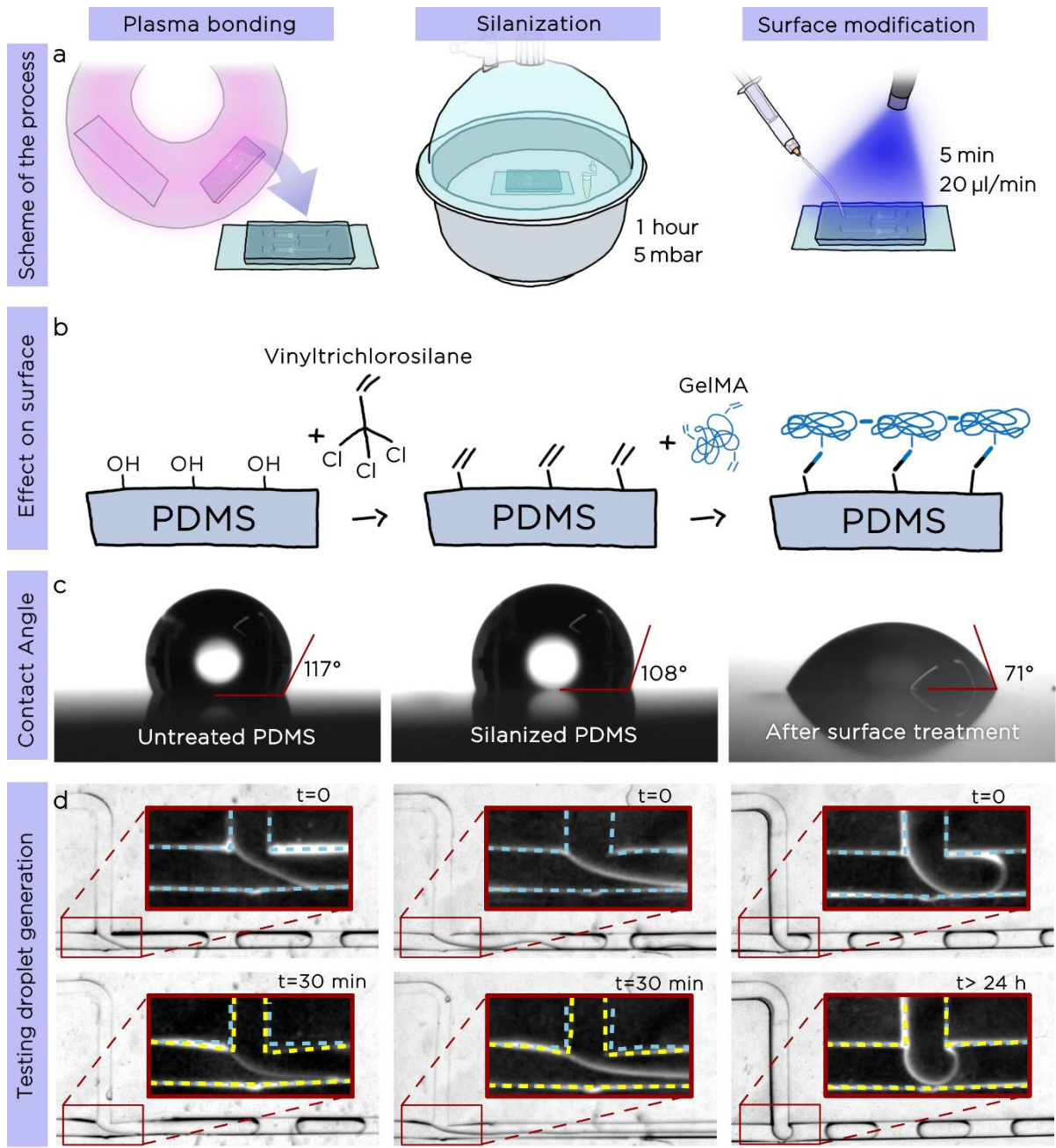

**Figure 3. Surface modification –** *a)* schemes of the steps for the surface modification (plasma bonding of the chip, silanization, and flushing GelMA solution while exposing the chip to UV light). *b)* Scheme of the effects on the surface after each step of surface modification. *c)* Measurements of contact angles on untreated PDMS, silanized PDMS and surface-modified PDMS. *d)* testing droplet generation. Only the treated PDMS surface remains unchanged for over 24 hours of constant droplet generation, while the other surfaces show swelling with hexadecane after just 30 minutes. We measured the width of the microchannel at the T-junction at the beginning of the experiment (t=0) and marked the channels with red dashed line. We then measured the microchannel at the same position and marked them in yellow for comparison after the noted time.

## 2.4    Characterization of Oil-in-Water emulsions in surface-modified Tuna-step

Since our ultimate goal was to use the Tuna-system as a printing head for the synthesis of porous hydrogels, we calibrated and characterized the surface-modified chips using as aqueous phases various biopolymer solutions, containing Dextran methacrylate (DexMA), Gelatin-methacryloyl (GelMA) or fibrinogen. For the disperse phase, we selected hexadecane. In these experiments, we fixed the flow rate of the aqueous phase at 1 µl/min and tuned the oil flow at 0.5, 1 and 2 µl/min, thus generating emulsions having droplet volume fractions of 33%, 50% and 66% respectively.

Similarly to the case of W/O emulsions, for each tested aqueous phase, the volume of the droplets reached the maximum value when the membrane was not actuated, while an increase in $P_m$ resulted in a decrease in nozzle and droplet sizes. We observed that the droplets are initially generated in the center of the nozzle, but as soon as we increase $P_m$, they give priority to one side of the actuated nozzle. In these cases, we noticed that the "preferred" side can be selected by slightly tilting the chip. When the pressure was increased further, the membrane effectively transformed the step geometry, creating two side nozzles (**Figure 4a**). Then, the droplets started being generated in parallel in the two side nozzles, until a critical pressure was applied to the membrane and the process of droplet generation became chaotic. For the tested aqueous phases, the dynamic ranges of the droplet sizes obtained were as follows (**Figure 4b**): i) the droplets in the fibrinogen solution ranged between 54 and 538 µm in diameter, corresponding to a volume range between 82 pl and 82 nl; ii) the droplets in the DexMA solution had a diameter ranging between 218 and 688 µm, corresponding to a volume range between 5 and 170 nl; iii) the droplets in the GelMA solution varied between 143 and 518 µm, corresponding to a volume range between 1.5 and 73 nl. The fibrinogen solution generated droplets in the widest pressure range, up to 360 mbar, and produced droplets in the widest range of volumes (3 orders of magnitude). However, we selected DexMA for the following experiments (i.e. the synthesis of our porous hydrogels) as, compared to the other two biopolymers, it can be easily and rapidly photopolymerized and the resulting gels offer better stability and mechanical response[25,26].

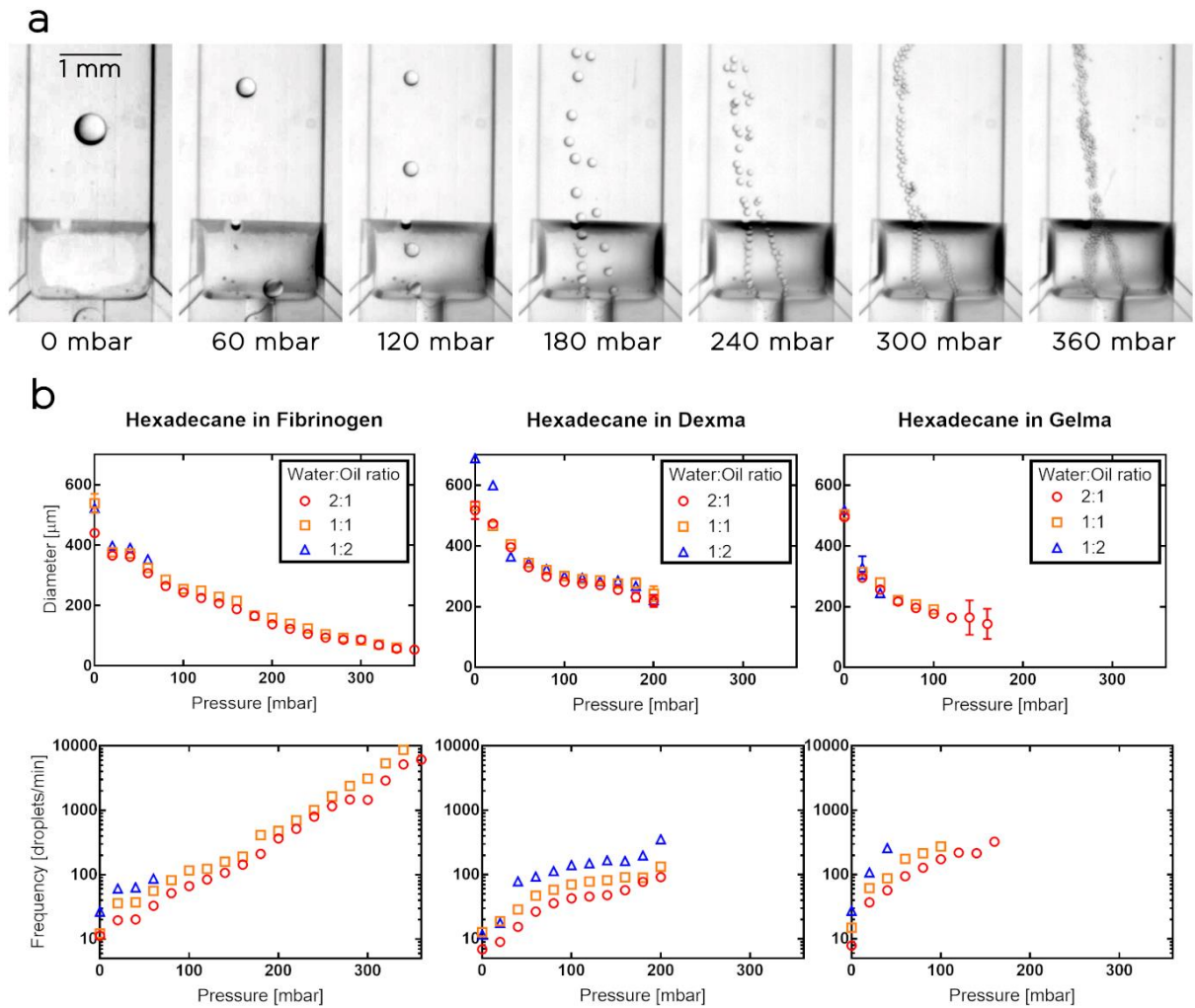

**Figure 4. Production of O/W emulsions within treated Tuna-step chip.** *a)* Generation of droplets of Hexadecane in Fibrinogen solution at increasing pressures on the membrane layer. Formation of two side nozzles from one initial nozzle visible for actuation pressures of 180 mbar and higher. *b)* Calibration graphs. Top graphs: diameter of the Hexadecane droplets in function of the pressure applied on the membrane layer (Aqueous phase from left to right: Fibrinogen, *DexMA* and *GelMA*). Error bars are displayed only when larger than the dots. Bottom graphs: frequency of droplet generation calculated dividing the flowrate by the volume of the corresponding droplets. Different colors represent the ratio between the aqueous and oil phases (Red – 2:1, Orange – 1:1, Blue – 1:2).

## 2.5 3D printing of porous functionally graded hydrogels (pFGhs) with Tuna-step

To finally 3D print porous functionally graded hydrogels, we mounted the Tuna-step chip on a custom 3D printer (**Figure 5a**). The printing process was exectued within an agarose fluid-gel bath that acted as a support material during emulsion extrusion. As demonstrated recently, the use of such fluid-gel bath is convenient as, from the one hand, allows the 3D printing of more complex architectures and, from the other one, it enables the decoupling of printing resolution from the rheological properties of the extruded emulsion[16] (**Figure 5b**).

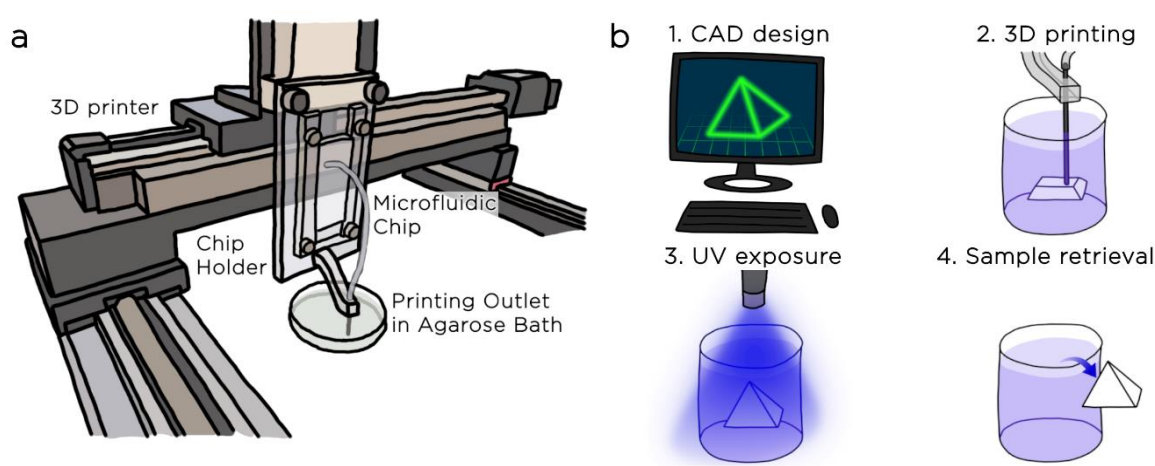

**Figure 5. 3D printing of pFGhs using Tuna-step chip.** *a)* Scheme of the 3D printing setup: the chip is mounted on an XYZ motorized gantry system. The outlet of the chip is connected to the printing nozzle that prints directly into the Agarose gel. *b)* Key steps to fabricate pFGhs: 1) CAD design (we design the object with CAD and translate the .stl file to a .gcode file to instruct the 3D printer); 2) 3D printing; 3) UV exposure to polymerize the external phase into solid hydrogel structure. 4) The 3D material is removed from the bath and processed through following steps of purification.

First, we tested our ability to extrude the O/W emulsion inside the agarose bath. To have a clear view over the position and size of the deposited droplets, we printed various 2.5D emulsion patterns using different pressure profiles for the membrane actuation (**Figure 6**). Moreover, we demonstrated that the pressure profile could be synchronized precisely with the robotic motion control, generating complex spatial emulsion patterns (see for instance the chess spatial organization of small and big droplets). **Figure 6c** and **d** demonstrate the synchronization of a square-function pressure profile with two distinct printing paths (serpentine and chess-like motion) with the scope of highlighting the wide array of spatial structures that can be achieved by finely adjusting the pressure profiles in relation to the spatial deposition.

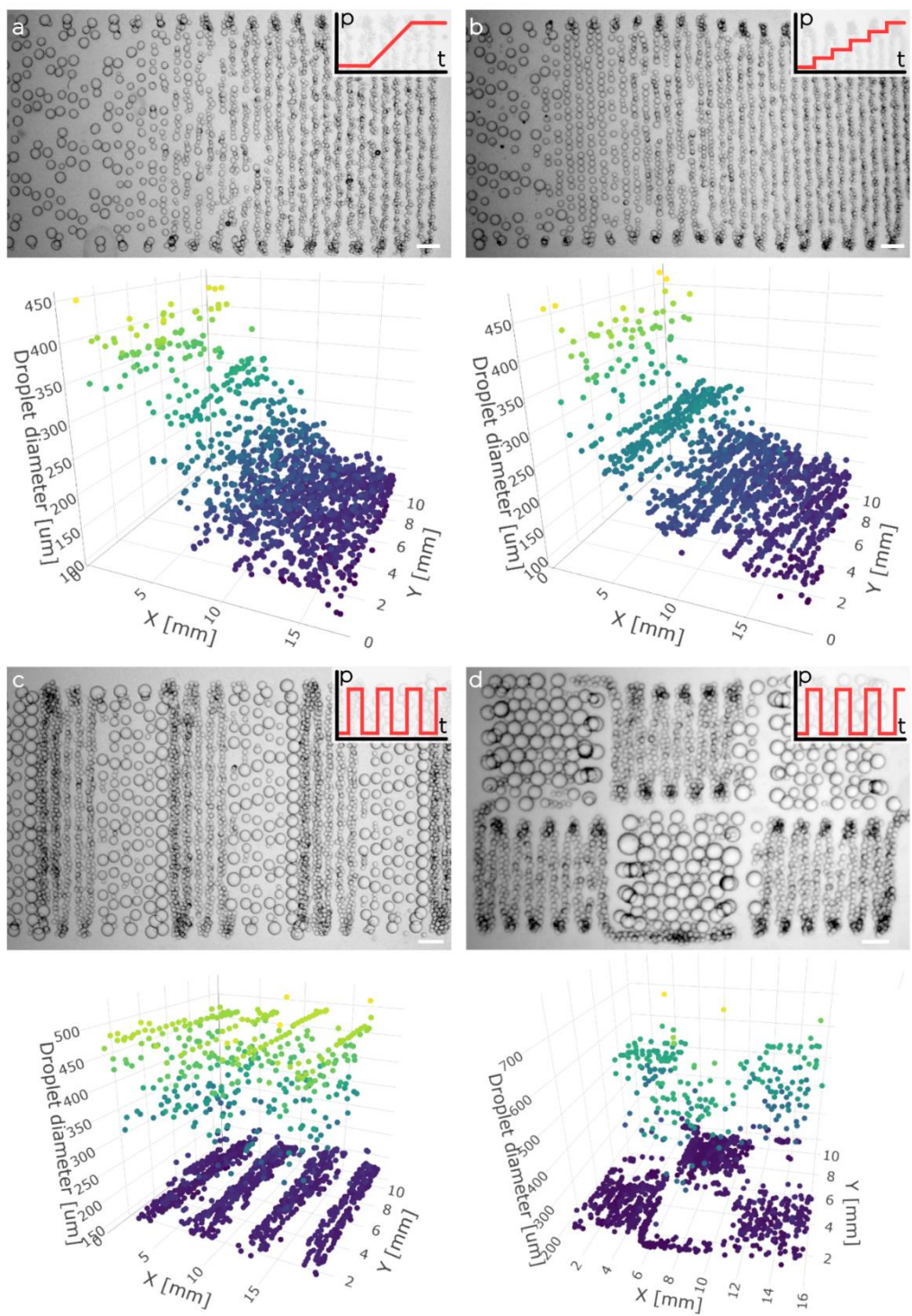

**Figure 6. 2.5D deposition of Hexadecane-in-DexMA emulsion in Agarose gel bath** – We actuated the PDMS membrane of the Tuna-step and consequently changed the droplet size. We coordinated the pressure gradients with the XY movement of the 3D printer, extruding the emulsion in the agarose bath and obtaining visual demonstrations of the validity of our method. *a)* and *b)* we applied an increasing pressure gradient, first gradually and then step-by-step. *c)* and *d)* we applied pressure following a square function, alternating 100 mbar to 0 mbar. The chess pattern obtained in the fourth panel is due to the path of the nozzle that covered the printing area with a square-by-square movement. Each picture is coupled with the graph showing the distribution on the XY plane of the droplet diameters measured. Scale bars correspond to 1 mm.

After achieving full synchronization between emulsion generation and spatial deposition within a layer, we tested our platform for the synthesis of macroscopic pFGhs. We selected six different geometries with various degrees of structural complexity and 3D printed them in the agarose support bath using, for convenience, an O/W emulsion having 20% of oil volume fraction. As shown in **Figure 7**, all structures were finely printed with minor discrepancies between the CAD model and the actual 3D shape.

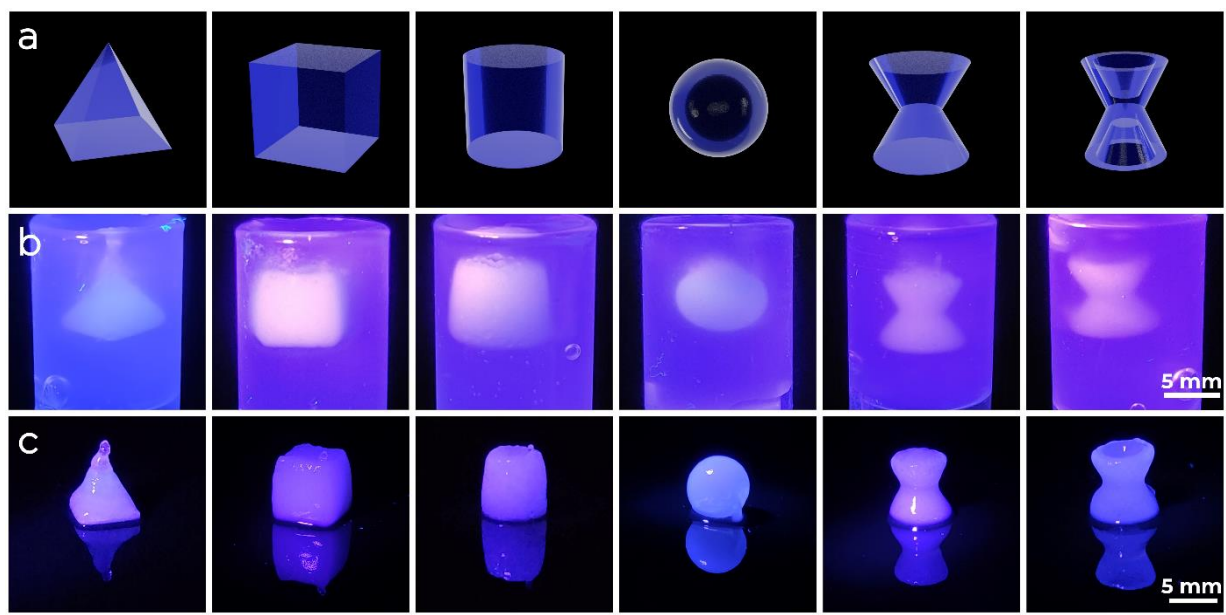

**Figure 7. 3D printed pFGhs structures made of DexMA.** *a)* 3D models of the expected structures. *b)* Completed structures in the agarose bath. *c)* Structures after the extraction, ready for further manipulation steps.

Finally, to additionally prove the possibility of printing multi-material O/W emulsions with Tuna-step, we added a Y-junction connector to the continuous phase inlet to process two different solutions. To better appreciate the spatial localization of the two materials within the printed O/W emulsions, we prepared two DexMA solutions containing a few mg/ml of fluorescent dyes. In this case, we supplied the two continuous phases to Tuna-step using different flow profiles and processed them in the form of complex 2.5D patterns (chess pawn, concentric circles and a mug of beer). The results are shown in **Figure 8**.

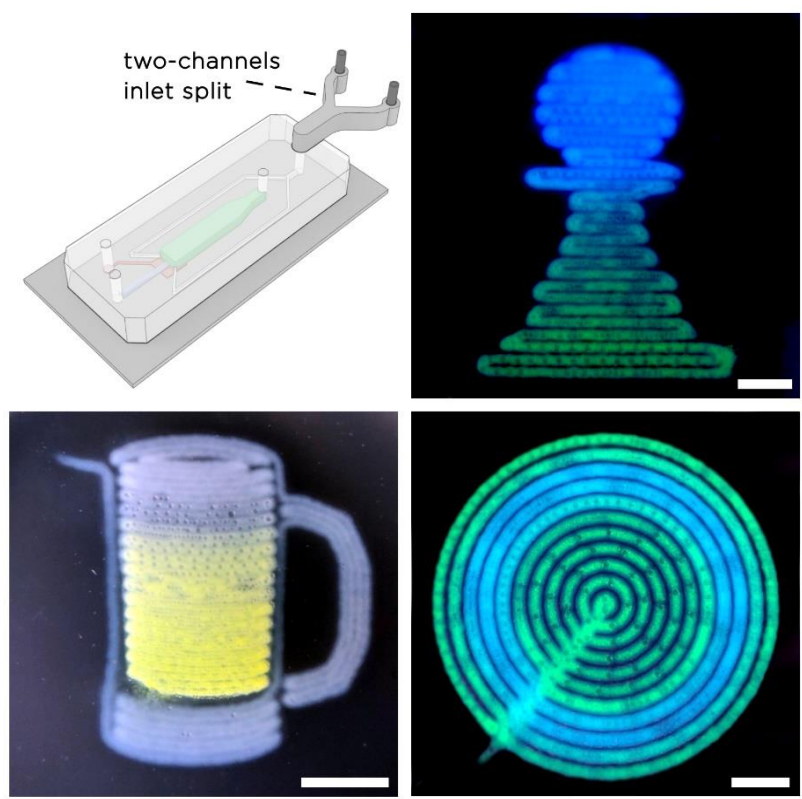

**Figure 8.** **Multi-material printing using Tuna-step.** Figures show three patterns to prove multi-material printing. We switched the emulsion outer phase during the extrusion and obtained the color change. Notice different colors (multimaterial printing) and different droplet sizes (printing materials of varying porosity) in all prints. Scale bars corresponding to 5 mm.

## 2.6 Parallelization of the Tuna-step nozzle to increase emulsion throughput

To show that Tuna-step is comparable in terms of throughput to flow focusing modules, we designed a microfluidic chip consisting of 14 parallel nozzles, with input from a single liquid source, aligned on top of a wide membrane (**Figure 9**). The inlet channel had a 400 µm height, and was connected to the nozzles (100 µm height) through linear slopes[27]. We incorporated the slopes to ensure a seamless transition from the inlet channel to the emulsifying nozzles while maintaining a significantly higher hydrodynamic resistance within the nozzles compared to the inlet channel. This design requirement is crucial to ensure a balanced flow rate for each nozzle, promoting uniformity in the emulsification process.

We calibrated and tested the multi-nozzle Tuna-step chip to generate water droplets in NOVEC 7500 oil (with 2% w/t surfactant). We applied a constant flowrate of 60 µl/min for the oil phase and 34 µl/min for the water (**Figure 9a**). We applied increasing pressures to the actuatable membrane with steps of 20 mbar, until we reach 200 mbar. After that, the droplet generation became unstable, so, based on the results obtained with single-nozzle chips, we decreased the flow rates to 14 µl/min for the oil and 8 µl/min for the water (**Figure 9b**); then, we increased again the pressure until we reach 300 mbar. We observed a gradual decrease in droplet size, with a range of diameters between 150±10 and 460±20 µm, corresponding to a volume range between 1.8±0.4 nl and 51±7 nl (**Figure 9c**).

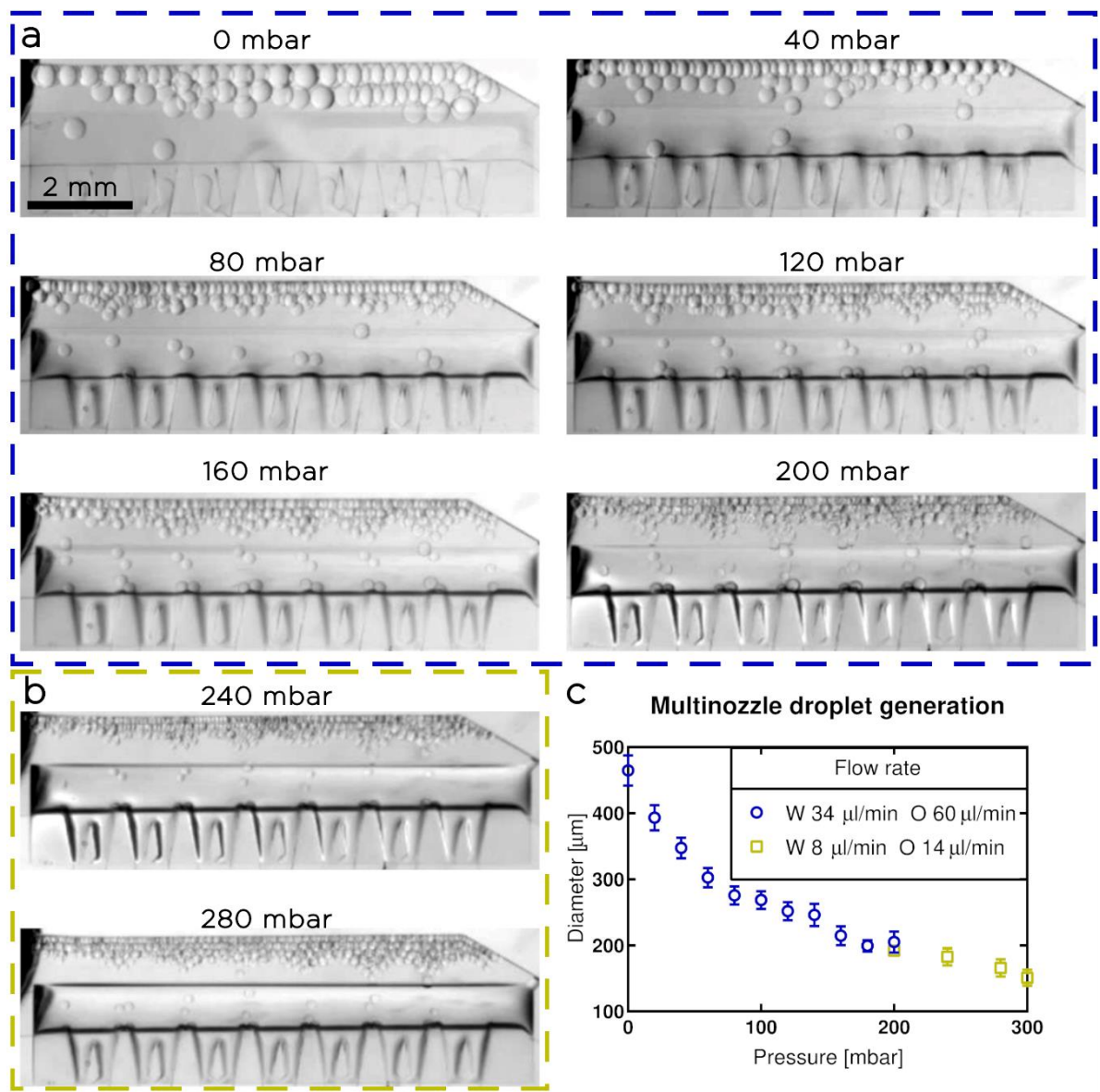

**Figure 9**. **Multi-nozzle generation of droplets at the Tuna-step**. Multi-nozzle Tuna-step during droplet generation operated at *a)* constant flow-rates (O = 60 µl/min, W = 34 µl/min) and increasing $P_m$, pressure up to 200 mbar, *b)* constant flow-rates (O = 14 µl/min, W = 8 µl/min) and $P_m$ between 200-300 mbar. *c)* Droplet diameter as a function of the pressure applied to the membrane. Oil phase: Novec 7500 with 2% w/t surfactant. Aqueous phase: distilled water.

## 3. Discussion

Here we reported Tuna-step, a droplet generation strategy based on a microfluidic step emulsification nozzle coupled with a flexible membrane for production of droplets of dynamically controllable sizes. By actuating the membrane beneath the step, we can control the nozzle lumen and thus control the size of generated droplets without changing the liquid input flow rates. Tuna-step combines advantages of flow focusing modules (controllable droplet volumes) and of step emulsification modules (constant input liquid flow rates, high volume fraction of obtained emulsions).

The working principle behind droplet generation in a step has been widely studied[5–7,10,27,28], and its correlation with the nozzle height is well known. In this configuration, the droplet size depends mostly on the nozzle size, resulting in the robust generation of highly monodisperse emulsions. However, when droplets of different sizes are necessary – as in the case of microfluidic-assisted 3D printing of pFGMs – conventional step emulsification systems are limited.

In the recent past, droplet size gradients in emulsions or foams were achieved in valve-based flow focusing systems[29,30]; however the strong dependence of flow-focusing geometries on the flow rate makes it challenging to parallelize them to increase the throughput. In contrast, the parallelization of step emulsification devices has been widely implemented for different high-throughput applications[7,9]. Here, we demonstrated that the Tuna-step geometry can be easily parallelized to generate droplets of controllable volume at a higher throughput, with potential for further upscaling. By multiplying the nozzles to 14, we generated W/O droplets at 130 Hz. Notably, the use of PDMS for chip manufacturing did not represent a major problem and both O/W and W/O emulsions could be generated in the proposed devices. We anticipate further studies that will parallelize the design we showed, exploiting the use of our technology for the high-throughput generation of droplets with a controllable size that could be implemented in other research areas beyond porous material synthesis. Based on our design, it is possible to engineer a highly parallelized chip for a high-throughput generation of monodisperse droplets with a diameter that is controlled only by the deflection and change of the geometry of the nozzle. Nevertheless, engineering such a design is not an obvious task. In fact, there are a few aspects that should be carefully addressed. First, it is necessary to ensure similar resistance through all the nozzles, otherwise the flow will prioritize those having less resistance, and the emulsification will not be consistent or parallelized. Second, a higher number of nozzles would require larger membranes, that can easily bond with the glass surface during chip manufacturing. This problem could be solved with the different design of the membrane or with a chip design that considers alternative materials, for example with an elastic thermoplastic membrane instead of a PDMS membrane. Third, the outlet of the chip should be designed carefully to facilitate the droplet escape from the nozzles. At high throughput and high rate of droplet generation, improper outlet design might trap the generated droplets in the chip, and thus destabilize droplet production in the long run.

Another characteristic aspect of our system is that the actuation of Tuna-step at high enough pressures forces formation of droplets in two places, effectively forming a double-nozzle from each Tuna-step nozzle. We hypothesize that there might be possibly an interplay between the two "sub-nozzles" during droplet generation; however, the potential differences in droplet sizes generated in the two sub-nozzles are not large enough to increase the overall emulsion polydispersity. We have reported before a step emulsification with a nozzle that was divided into two sub-nozzles by a solid obstacle in the middle of the nozzle[31]. Such geometry generated droplets for different flow rates keeping the size constant, and the design was considerably different from the one reported in this work. In this paper, even when the membrane acts as an obstacle in the middle of the nozzle, it constitutes a flexible part that is changing dynamically in function of the overall pressure. The result is a series of monodisperse droplets with a size that depends on the pressure on the membrane layer. As reported by Unger et al.[20], the response of the valve to pressure actuation can be adjusted by changing the membrane thickness. In our study, we used a membrane with a thickness of ~50 µm, allowing a reasonable pressure range in relation to the membrane deflection. A thinner membrane would be deflected with a lower pressure range, obtaining a coarse tuning in size. A thicker membrane would require higher pressures for actuation, that could damage the chip or temporarily deform the structures in the thick layer (we already observed that the nozzles in the thick layer are subject to deformation when we applied higher pressures). In the future, one could easily tune the Young modulus of PDMS to obtain more or less flexible structures, by changing the ratio between the polymer and the curing agent, or could adopt different materials. With PDMS being one of the widest adopted materials in microfluidics, the PDMS-based technology that we developed should be easily accessible to other researchers.

In this study, we aimed at using the developed Tuna-step in combination with 3D printing to manufacture porous biocompatible hydrogels. In this case, the use of a hydrophilic continuous phase required the PDMS microfluidic chip to be hydrophilic. To accomplish this, we developed a novel hydrophilic surface modification. Notably, our surface treatment was particularly stable and durable, enabling us to use the same chip several times without repeating the modification. We coupled our chip with a custom 3D printer, and we deposited the emulsion in a supporting agarose bath. During the emulsion deposition, we applied different pressure profiles to the membrane, generating tunable emulsions. By synchronizing the mechanical motion with the emulsion production, we showed that one can create sophisticated 2.5D and 3D droplet patterns which after the crosslinking of the continuous phase, will represent the porous template. Of note, we adapted the inlet of the chip to use two different materials for the outer phase, showing evidence of multi-material printing of pFGhs. In the future, some efforts should be spent to further optimize the tuna-step design and fabricate pFGhs within even wider operational ranges. In particular, the control over the volume fraction of the dispersed phase should be improved. This parameter, in fact, is primary responsible for the porosity of the fabricated material, a feature that is often desirable to tune in wide ranges to tailor, for instance, the sample mechanical properties. At the moment, our design operates properly up to a droplet volume fraction of 60%. Above this value, the steps are crowded with droplets and droplet generation becomes inconsistent. A possible solution that could be adopted consists in the use of draining channels in the chip outlet for partially removing the excess of continuous phase[32,33].

We showed that Tuna-step is a valid droplet generation strategy, and that it could benefit from some developments in the future. The most important are large-scale parallelization of nozzles, and development of Tuna-step in materials that are easier to produce at mass scale than PDMS, like thermoplastics.

## 4. Conclusions

In this study, we introduced Tuna-step, an innovative droplet microfluidic system that overcomes a key limitation of step emulsification: fixed droplet size. Utilizing Tuna-step, we successfully generated droplets of varying sizes and compositions using a step-emulsifying nozzle. Our dynamic step emulsifier enabled us to produce emulsions with gradient droplet sizes. We extensively characterized our device at different flow rates to determine the operational range for oil-in-water and water-in-oil emulsions. By adjusting the pressure on the Tuna-step membrane, we achieved consistent reduction in droplet size across the tested flow rates. To facilitate the generation of oil-in-water emulsions, we developed a custom hydrophilic surface modification that prevented PDMS swelling with hexadecane oil. With this modification, we were able to continuously produce oil-in-water droplets for over 24 hours. We leveraged our technology to 3D print functionally graded materials, incorporating changes in porosity and material composition. By placing the chip on a custom 3D printer, extruding the emulsion into an agarose gel support bath, and exposing it to UV light for polymerization, we obtained self-sustaining 3D structures after removing the porous material from the bath. Our study not only demonstrates the suitability of our Tuna-step design for 3D emulsion printing and materials science but also opens up possibilities for a wide range of future applications. This consideration is supported by the preliminary results obtained in the parallelization of the Tuna-step geometry, where we produced custom emulsions from 14-nozzles working simultaneously.

Looking ahead, we envision further parallelization of the presented systems, enabling high-throughput emulsion generation with active control over droplet size, regardless of flow rate or volume fraction.

## 5. Materials and methods

### 5.1 Design of Tuna-step microfluidic chip

We realized a two-layer PDMS chip replicating the on-chip valve technology[20] and adapted the design to overlap the thin PDMS membrane with the nozzle for step emulsification, as shown in **Figure 1**.

The chip consists of two PDMS layers bonded on a 1 mm thick glass slide. The layer in contact with the glass slide is thin (100 µm), and it includes the chips membrane and one inlet to access the source for pressurized air. The layer on top is thick (5 mm), and it includes one inlet for the disperse phase, one for the continuous phase, and one outlet for the definitive collection of the droplets.

### 5.2 Master preparation

We realized the scheme of the chip using CAD (Autodesk AutoCAD) and used different approaches to produce the masters for the two separate layers. 1) *Thin layer*: we prepared the master with standard photolithography[34]. We coated a 7" silicon wafer with a few ml of SU8-2050 and performed the spin-coating for 30 s at 3000 rpm. We soft-baked the chip for 1 minute at 65 °C and then 6 minutes at 95 °C. We exposed the chip for 19 s and performed the post-exposure baking (1 minute at 65 °C and then 5 minutes at 95 °C). We immersed the wafer in SU8-developer for 5 minutes to remove the unexposed resist and rinsed the chip with isopropanol. We air-dried the chip and performed the hard baking at 200 °C for 5 minutes, and then we let the wafer cool down to room temperature. We silanized the wafer to prevent PDMS adhesion: we put the wafer in a vacuum chamber with two test tubes containing 30 µl of silane (Trichloro (1H, 1H, 2H, 2H-perfluorooctyl) Silane, Sigma Aldrich, USA), and activated the pump for 45 minutes at 10 mbar.

### 5.3 Chip preparation

1) *Thin layer:* We deposed the thin layer with a spin-coater: we covered the center of the chip with a few drops of uncured PDMS (10:1 ratio between the PDMS and the crosslinking agent). We controlled the thin layer thickness by adjusting the speed of the spin-coater (1500 rpm for 30 s) and the time after mixing PDMS with its curing agent (3h waiting time). We oven-baked the thin layer for 2 hours at 75 °C. We obtained thin PDMS layers with a ~100 µm thickness. The channel depth in the thin layer was ~50 µm, hence the thin membrane below the step had thickness ~50 µm. *2) Thick layer*: we prepared the master by CNC milling on 5 mm polycarbonate plates. We replicated the chip by coating the milled polycarbonate plates with PDMS and baking it for 2 hours at 75 °C. We then silanized the negative mold to prevent adhesion and repeated the procedure to obtain the positive replica. We baked it for 2 hours at 75 °C and finally removed the thick layer from the mold. We punched the inlets and outlets for the flow layer. *3) PDMS on PDMS bonding*: we poured a few drops of curing agent on a clean silicon wafer and spin-coated it at 5500 rpm for 45 s to obtain a thin, uniform coating. We positioned the thick PDMS layer on the wafer, ensuring that the surface was lightly coated with the curing agent. Following the markers on the thin layer, we placed a homemade frame on the thin layer to align the two layers, and then we inserted the thick layer within the frame, carefully removing the bubbles. We baked the chip overnight at 75 °C, then slowly peeled away the thin layer from the silicon master, leaving two bonded PDMS layers. We punched the hole for the thin membrane inlet and sealed the chip with the glass slide via standard plasma bonding. After bonding, we baked the chip for 10 minutes at 75 °C to help the adhesion of the PDMS on the glass.

### 5.4 Hydrophobic surface modification

Following the bonding process, we flushed the channels with a 5% w/t solution of silane (Trichloro (1H, 1H, 2H, 2H-perfluorooctyl) Silane, Sigma Aldrich, USA), in fluorinated oil (NOVEC 7500, 3M, USA). We carefully avoided applying excessive pressure that could result in the membrane coming into contact with the glass slide, rendering it unusable.

### 5.5 Water-in-oil droplet generation

We filled one glass syringe (Hamilton gastight) with water and one with oil (Novec 7500 oil with 2% w/t PFPE-PEG-PFPE surfactant, synthetized according to the protocol described by *Holtze et al.*[35]). Each syringe is connected to a 0.5 mm O.D. steel needle and then to the inlet of the chip through a Teflon tubing (0.8 mm O.D., 0.5 mm I.D.). We controlled the flowrates with precision syringe pumps (neMESYS, Cetoni) with the related software (neMESYS user interface).

We controlled the pressure in the thin layer with a pressure controller (OB1-MK3+, Elveflow), generating stationary pressure gradients or alternated pressure curves with the related software.

### 5.6  Hydrophilic surface modification

Immediately after the bonding, we put the chip in the vacuum chamber with a test tube containing 30 ml of Vinyltrichlorosilane. We activated the vacuum pump for 1 hour at 5 mbar. We removed the chip from the vacuum chamber and flushed the channels with a 1% w/t solution of Gelatin-methacryloyl (GelMA) in water at a constant flow rate of 20 µl/min. During this process, we shine UV light on the chip, activating the bonding between GelMA polymers and the previously introduced Vinyltrichlorosilane. Finally, we rinsed the chip with distilled water to remove the GelMA solution.

### 5.7  Generation of Oil-in-Water droplets

We generated droplets of Hexadecane in three different aqueous solutions. We prepared new batches of aqueous solutions shortly before every experiment to avoid contamination of the reagents or photoinitiation of the DexMA solution. *GelMA solution:* We prepared a 0.5% w/t solution of Plantacare in water and then solved GelMA to form a 3% w/v solution. *DexMA solution:* We prepared a 0.5% w/t solution of Plantacare in water, and then solved DexMA to form a 20% w/v solution. We prepared a 100 mg/ml solution of Irgacure in a 70% w/t solution of Ethanol in water, that we used as initiator with a 100x dilution. *Fibrinogen solution:* We prepared a 1.4% w/t solution of fibrinogen in 0.5% w/t solution of Plantacare in water.

### 5.8  3D printing in Agarose fluid gel

We prepared the bath of Agarose fluid gel adding 0.5% w/t agarose in water. We autoclaved the solution at 121 °C for 20 minutes to sterilize and melt the agarose solution, and then we stirred the solution overnight at 700 RPM. With the gradual decrease in temperature, agarose gel microparticles form, and at room temperature, we obtain an agarose fluid gel (as previously reported[36]).

We stored the oil and aqueous phases in glass syringes and controlled the flow with syringe pumps, as described before.

We secured the chip to the z-axis of the XYZ robotic gantry system in a vertical position to favour droplet escape from the step nozzle using a a custom-made polycarbonate holder that was equipped with a 25G metallic needle for precise emulsion deposition (**Figure 5a**).

The printing process consists of five steps: i) We design the 3D model of the object with computer-aided design (CAD, Autodesk Inventor) ii) We use free software (Slic3r) to convert the CAD file into a file containing the instructions for the XYZ movements of the printer stage (G-code file). We performed the automated slicing of the 3D object considering a printing nozzle with a 0.5 mm diameter and a slice thickness of 0.2 mm. iii) We generated an emulsion of Hexadecane droplets in DexMA solution and extruded the emulsion in the agarose bath. During this step, the motorized stage follows the instructions of the G-code file, depositing the emulsion within the agarose bath (**Figure 6a**). iv) Following the extrusion, we exposed the emulsion to UV light (λ = 365 nm) for 5 minutes, triggering the crosslinking of the material (**Figure 6b**). v) After the polymerization, we extracted the 3D structures from the bath, rinsed them with water, and proved their stability (**Figure 6c**).

**Conflicts of interests**

There are no conflicts to declare.

**Acknowledgements**

This article is part of a project that has received funding from the European Union's Horizon 2020 research and innovation programme under the Marie Skłodowska-Curie grant agreement No 813786 (EVOdrops). M.C. and M.C.T. acknowledge the funding from the National Science Centre Poland (NCN) within OPUS 19 project No. 2020/37/B/ST8/02167. W. P. was supported by the Polish National Agency for Academic Exchange (NAWA) through the Bekker Programme, grant no. PPN/BEK/2020/1/00333/U/00001. W. P. was supported by the Foundation for Polish Science (FNP) with the START 069.2021 scholarship. P.G. acknowledges the support from the National Science Centre, Poland, funding based on decision 2018/30/A/ST4/00036, Maestro 10.

**Authors' contribution**

**Francesco Nalin**: Conceptualization, Data Curation, Formal Analysis, Investigation, Methodology, Visualization, Original Draft Preparation, Review & Editing. **Maria Celeste Tirelli**: Investigation, Review & Editing. **Piotr Garstecki**: Funding Acquisition, Project Administration, Resources, Supervision, Review & Editing. **Witold Postek**: Conceptualization, Supervision, Review & Editing. **Marco Costantini**: Conceptualization, Funding Acquisition, Methodology, Project Administration, Resources, Supervision, Visualization, Review & Editing.